\documentclass[sigconf, anonymous=false, review=false]{acmart}
\usepackage{algorithm}
\usepackage{algorithmic}
\usepackage{balance}
\usepackage{multirow}
\usepackage{float}

\AtBeginDocument{%
  }

\setcopyright{cc}
\setcctype{by}
\copyrightyear{2018}
\acmYear{2018}
\acmConference[Conference acronym 'XX]{Make sure to enter the correct
  conference title from your rights confirmation email}{June 03--05,
  2018}{Woodstock, NY}
\acmISBN{978-1-4503-XXXX-X/2018/06}
\acmDOI{XXXXXXX.XXXXXXX}

\begin{document}

\title{Don't Rewrite the SID Map: Topology Preservation for Generative Retrieval in Dynamic E-Commerce Search}


\author{Jianbo Zhu}
\correspondingauthor
\affiliation{
  \institution{Taobao \& Tmall Group of Alibaba}
  \city{Hangzhou}
  \country{China}
}
\email{zhujianbo.zjb@taobao.com}

\author{Xing Fang}
\affiliation{
  \institution{Nankai University}
  \city{Tianjin}
  \country{China}
  }
\affiliation{
  \institution{Taobao \& Tmall Group of Alibaba}
  \city{Hangzhou}
  \country{China}
}
\email{fangxing.fx@taobao.com}

\author{Xiaopeng Chu}
\affiliation{
  \institution{University of Science and Technology of China}
  \city{Hefei}
  \country{China}
}
\email{cxphhh@mail.ustc.edu.cn}

\author{Mingmin Jin}
\affiliation{
  \institution{Taobao \& Tmall Group of Alibaba}
  \city{Hangzhou}
  \country{China}
}
\email{jimmy.jmm@taobao.com}

\author{Jing Wang}
\affiliation{
  \institution{Taobao \& Tmall Group of Alibaba}
  \city{Hangzhou}
  \country{China}
}
\email{jing.wangj1@taobao.com}

\author{Bokang Wang}
\affiliation{
  \institution{Taobao \& Tmall Group of Alibaba}
  \city{Hangzhou}
  \country{China}
}
\email{wangbokang.wbk@taobao.com}

\author{Jiaqi Ye}
\affiliation{
  \institution{Taobao \& Tmall Group of Alibaba}
  \city{Hangzhou}
  \country{China}
}
\email{guike.yjq@taobao.com}

\author{Xiang Zhang}
\affiliation{
  \institution{Taobao \& Tmall Group of Alibaba}
  \city{Hangzhou}
  \country{China}
}
\email{heiqiang.zx@taobao.com}

\renewcommand{\shortauthors}{Jianbo Zhu et al.}

\begin{abstract}
Generative retrieval assigns each item a discrete semantic identifier (SID) and retrieves items by autoregressively generating SID sequences. In dynamic environments such as e-commerce, new items arrive continuously and business signals keep evolving, requiring periodic updates to the SID-to-Item mapping and model to maintain retrieval timeliness. Each mapping update, however, can invalidate search paths already learned by the retriever model, and the quantitative relationship between the degree of mapping change and retrieval quality has not been systematically characterized. We study this problem with extensive controlled experiments over 67 days of large-scale e-commerce search logs and obtain three core findings. (1)~SID semantic drift causes severe retrieval degradation that may be masked by large beam sizes: mapping updates reduce HitRate by 41.78-62.05\% at beam size~1 but only 8.33-10.85\% at beam size~1000, as redundant candidate generation conceals the true severity of drift. (2)~The fraction of changed SIDs is strictly monotonically related to retrieval quality: as the remapped fraction of old-item SIDs increases from 0\% to 72.75\%, HitRate decreases monotonically, with a drop of up to 42.66\% at beam size~1. Under the same warm-start training procedure, full remapping remains substantially worse than the preservation policy, indicating that the model has internalized the implicit topology of the SID space and that short-budget continued training alone cannot repair topology disruption. (3)~The SID topology-preserving principle makes even the simplest warm-start strategy highly effective, and models fine-tuned from previous checkpoints outperform naive model souping: fixing old-item SIDs, inserting new-items, and warm-starting from the previous model improves old-item HitRate by approximately 10\% and new-item HitRate by approximately 109\%, validated across two non-overlapping update cycles. These findings distill into a concise deployment principle: insert new-items, freeze old-items; warm-start when possible, otherwise freeze; never change SIDs the model has already learned. A three-week online A/B test of this principle on the mobile e-commerce platform TmallAPP shows UCTCVR +0.29\% and GMV +1.16\%.
\end{abstract}

\ccsdesc[500]{Information systems~Information retrieval}
\ccsdesc[500]{Information systems~Retrieval models and ranking}

\keywords{Generative Retrieval, Semantic Identifiers, E-Commerce Search, Semantic Drift, Dynamic Update}


\maketitle

\section{Introduction}
Generative retrieval reformulates item retrieval as a sequence generation task: given a user query, a language model autoregressively generates a short discrete token sequence, known as a semantic identifier (SID), which is resolved to one or more items through an external mapping. Since the differentiable search index~\cite{tay2022transformer} was proposed, this paradigm has demonstrated advantages in many scenarios~\cite{decao2021genre,rajput2024tiger,damico2025spotify,deng2025onerec,chen2025onesearch,zhu2026ecommerce}, offering end-to-end training, elimination of large-scale vector index maintenance, and natural integration of multi-objective ranking signals.

Beyond the initial item SID construction, a more pressing challenge in dynamic settings is how to maintain the SID-to-Item mapping over time. In e-commerce search, a highly dynamic retrieval environment and the item set expands daily: new items are listed, prices fluctuate, and conversion rates evolve, all of which require periodic updates to the retriever model and SID-to-Item mapping. In production-grade systems in particular, SID encodings often incorporate a business-value dimension~\cite{ling2026businessvalue}, making the mapping unavoidably time-varying. After the model is trained on a historical mapping, it may generate the same SID at inference time, yet that SID now resolves to a different item under the updated mapping---a phenomenon we call \emph{semantic drift}. A deeper issue is that prior work has predominantly pursued collision-free SID assignments; in dynamic settings, however, the continuously growing item set inevitably introduces collisions with existing items. Forcing a full remapping to eliminate these collisions changes the SIDs of many existing items, triggering severe semantic drift.

In practice, semantic drift is easy to overlook. At a large beam size, the model can still recover target items by exploring many alternative branches even when high-probability paths have become stale; offline evaluation at a single wide beam may therefore suggest that a mapping update is safe, while latency-constrained serving at a narrow beam suffers a far larger loss. Meanwhile, retraining is often assumed to repair the refreshed mapping. However, this assumption neglects the structural knowledge the model has already encoded during training, particularly the learned topology of the SID space.

These considerations raise some deployment-oriented questions. How much damage does a SID mapping refresh cause, and how does the observed damage depend on beam size? Does retrieval quality vary systematically with the fraction of old items whose SIDs change? Can warm-start training repair the damage from remapping, or should the system preserve the old SID topology and only insert new items?

Rather than proposing another algorithmic remedy, we take a step back and answer these questions empirically through extensive controlled experiments over 67 days of large-scale e-commerce search logs. Unlike prior studies that evaluate a single proposed method against baselines, we systematically vary the mapping policy as an experimental factor, spanning the full spectrum from complete remapping to complete freezing. By independently controlling the training strategy, this design enables a clean disentanglement of mapping-induced damage from training-induced repair. Our contributions are:

\begin{itemize}
  \item \textbf{Damage and masking effect of semantic drift.} We systematically quantify the degradation caused by SID mapping evolution and reveal that wide-beam evaluation masks the true severity of drift. Specifically, while mapping updates cause a notable reduction in HitRate and MRR at beam size~1000, the degradation is substantially more severe at beam size~1.
  \item \textbf{Monotonic effect and topology irreparability.} We discover a strict monotonic negative relationship between the fraction of changed old-item SIDs~$\gamma$ and retrieval quality. Under the same warm-start budget, full remapping remains substantially worse than the preservation policy, indicating that the model has internalized the implicit topology of the SID space and that the simple warm-start regime does not repair topology disruption.
  \item \textbf{Topology-preservation principle and optimal update policy.} We find that topology preservation is a more fundamental determinant than retraining alone, and propose a simple incremental warm-start strategy feasible at industrial scale (freeze old items, insert new items, warm-start the model periodically). Experiments show approximately 9\% improvement on old items and 109\% on new items; a three-week online A/B test yields UCTCVR +0.29\% and GMV +1.16\%.
\end{itemize}

Four findings are particularly counter-intuitive. First, evaluations conducted with wide beams can mask the vast majority of the actual drift severity, meaning that many production systems may be operating under a false sense of safety. Second, the relationship between remapping fraction and quality loss is strictly monotonic, leaving no useful middle ground between full remapping and complete freezing. Third, warm-start training, the default repair strategy assumed by most continual-learning approaches, cannot recover from topology disruption: Full-Remapping after warm-start still underperforms a no-retrain Frozen baseline. Fourth, rather than creating a 'model soup' by mixing the old and new models, it is more effective to directly use the new model trained from the old one. Together, these findings suggest that the problem prior work has been solving, namely how to repair mapping-induced damage, is better addressed by not creating the damage in the first place.

\section{Related Work}
\subsection{Generative Retrieval}
Differentiable Search Index~\cite{tay2022transformer} and GENRE~\cite{decao2021genre} established the paradigm of directly generating document or entity identifiers, replacing nearest-neighbor search with autoregressive sequence generation. Since then, generative retrieval has been applied across document and passage retrieval~\cite{tay2022transformer,decao2021genre}, podcast discovery~\cite{damico2025spotify}, recommendation ranking~\cite{rajput2024tiger,singh2024semanticids,deng2025onerec}, and e-commerce search~\cite{chen2025onesearch,zhu2026ecommerce,ling2026businessvalue,zhan2026tsgrtaobaosearchgenerative}. TIGER~\cite{rajput2024tiger} demonstrated that residual-quantized semantic identifiers can effectively capture item semantics for retrieval, and since its introduction, a wave of work has adopted semantic IDs in e-commerce. OneRec~\cite{deng2025onerec} and OneSearch~\cite{chen2025onesearch} propose an end-to-end framework. In e-commerce search scenarios, CQ-SID~\cite{zhu2026ecommerce} introduces query-item contrastive learning and EG-GRPO for search-relevant clusters. Ling et al.~\cite{ling2026businessvalue} encode business-value ranking into SID assignments, and Zhan et al.~\cite{zhan2026tsgrtaobaosearchgenerative} propose a unified generative framework that integrates retrieval and pre-ranking within a single model. These efforts focus on initial model design; the update-time implications of evolving item collections remain open.

\subsection{Semantic Identifier Construction}
The quality of semantic identifiers is a first-order determinant of generative retrieval performance. RQ-VAE-based methods~\cite{rajput2024tiger,chen2025onesearch,zhu2026ecommerce} produce hierarchical codes through residual vector quantization of item embeddings. Alternative approaches include RQ-KMeans~\cite{deng2025onerec}, multi-identifier pretraining~\cite{zheng2025mtgrec}, and self-improving tokenization~\cite{chen2024selfimproving}. A key concern is how quantization distorts neighborhood structure: topology-aware tokenization~\cite{liu2026topology}, qualification-aware collision handling~\cite{hu2025quasid}, and topology-preserving tokenization~\cite{wang2026topogr} all indicate that identifier structure affects downstream retrieval. Hybrid semantic/hash identifiers~\cite{dong2025h2rec} and constrained decoding~\cite{tang2026static} address serving efficiency, while FORGE~\cite{fu2025forge} provides a systematic benchmark for evaluating SID construction methods. DREAM~\cite{guan2026dream} further identifies premature static SID commitment as a cold-start bottleneck and proposes counterfactual contrastive learning for progressive refinement. These efforts focus on constructing higher-quality initial SIDs; however, even when the initial design is effective, the impact of subsequent mapping evolution on retrieval quality remains insufficiently characterized.

\subsection{Dynamic Updates and Continual Learning}
DSI++~\cite{mehta2023dsiplus} frames document addition as continual learning and documents catastrophic forgetting during sequential updates. CLEVER~\cite{chen2023clever} and PromptDSI~\cite{huynh2024promptdsi} address dynamic corpora under limited replay. More recent work directly targets semantic ID staleness: Baikalov et al.~\cite{baikalov2026staleness} propose bijective alignment mappings to mitigate collaborative-SID drift, Feng et al.~\cite{feng2026dact} introduce drift-confidence-based selective tokenization to minimize unnecessary code changes. Alternative paradigms avoid weight updates entirely: Den et al.~\cite{den2026icicle} reframe incremental retrieval as an in-context generation problem, Mekonnen et al.~\cite{mekonnen2026pamt} attach parametric memory heads to frozen models, and Huynh et al.~\cite{huynh2025mixlora} use dynamically expandable LoRA experts. These methods mitigate mapping staleness through increasingly sophisticated algorithmic designs such as alignment solvers, selective reassignment, and parameter-efficient adapters. However, most of these approaches introduce additional architectural complexity.

Although recent methods have recognized the issue of drift and proposed algorithmic remedies, no study has treated the mapping policy as an independent experimental factor for disentangling the damage caused by mapping changes themselves from the effectiveness of repair algorithms. Nor has the monotonic relationship between the fraction of changed mappings and retrieval quality been established. More fundamentally, a simpler yet prerequisite question, whether the mapping should be changed at all, has been largely overlooked.

\begin{figure*}[t]
\centering
\includegraphics[width=\textwidth]{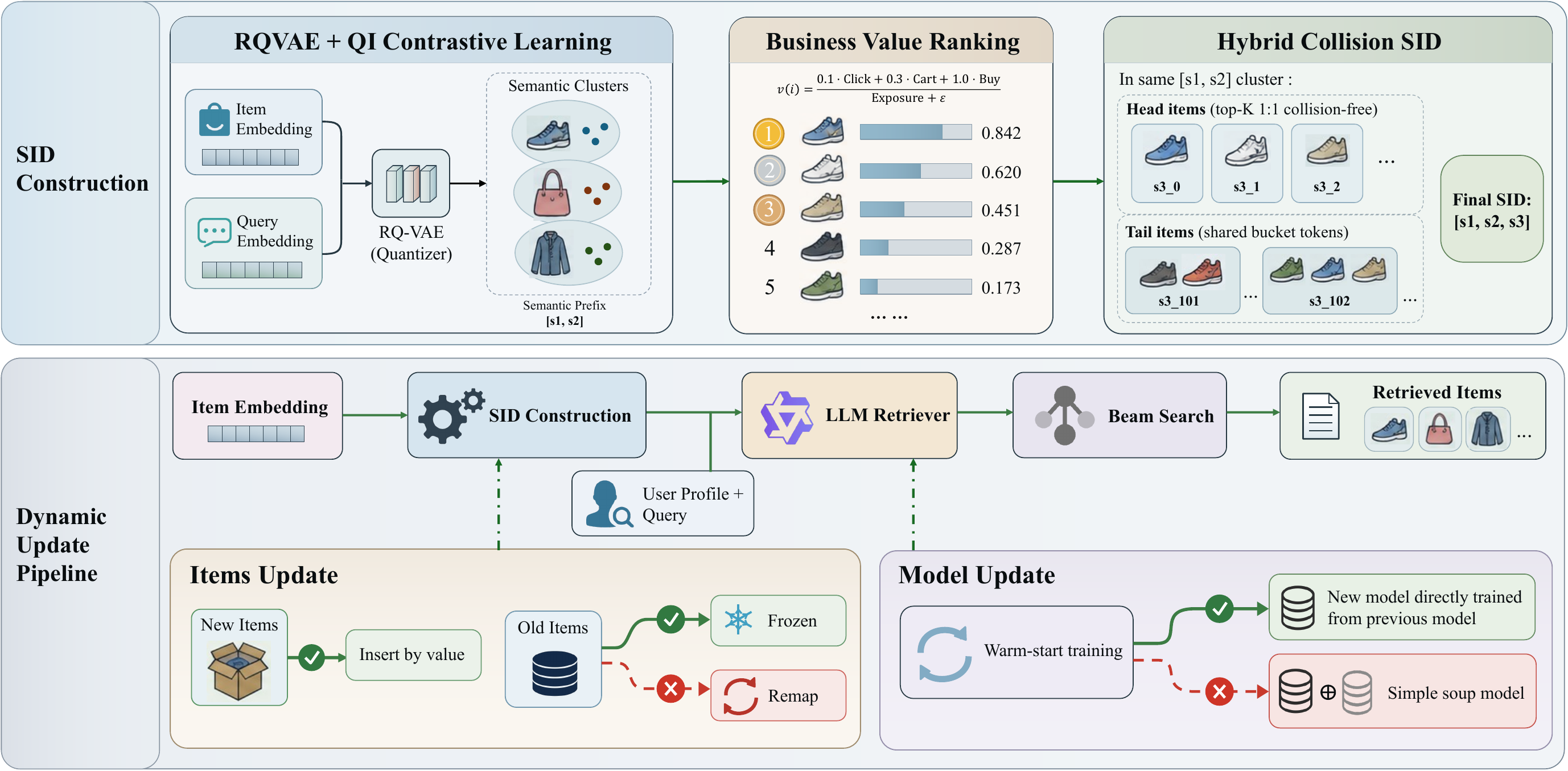}
\caption{\textbf{Framework overview.} \textbf{Top:} Three-level SID construction, two semantic layers from RQ-VAE with QI contrastive learning produce prefix $(s_1, s_2)$; a business-value layer appends $s_3$ by ranking items with engagement signals. \textbf{Bottom:} Retrieval pipeline from embedding through SID construction, LLM autoregressive generation, beam search, and SID-to-item map resolution. There are two divergent types of update paths: the red paths rewrite old item SIDs and use a soup model (combining old and new models), which degrades metrics; the green paths freeze old SIDs, insert new items, and directly apply the new model fine-tuned from the old one, improving performance on both old and new items.}
\label{fig:system}
\end{figure*}

\section{Method}
We study how to update a generative retrieval system deployed in e-commerce search when the item set expands and business-value signals shift. During long-term operation in real-world environments, the deployed retriever gradually becomes outdated as new items arrive and the commercial importance of existing items changes. Directly refreshing the Item-SID mapping to catch up with changes will destroy the autoregressive decoding path learned by the model, resulting in the decline of retrieval quality. Our goal is to identify update policies that accommodate new items and evolving value signals while keeping the learned mappings for existing items as stable as possible, avoiding a full rebuild-and-retrain pipeline.

\subsection{Problem Formulation}
Let $\mathcal{I}$ denote the item set. Each item $i\in\mathcal{I}$ is assigned a three-token semantic identifier
\begin{equation}
\operatorname{SID}(i)=[s_1,s_2,s_3].
\end{equation}
Given the query $q$ and user profile $u$, the retrieval model $f_\theta$ generates the SID of the target item autoregressively and is trained by minimizing
\begin{equation}
\mathcal{L}(q,u,i)=-\sum_{k=1}^{3}\log P_\theta(s_k\mid s_{<k},q,u).
\end{equation}
At inference, beam search produces the top-K SID sequences, and a serving-time mapping $M:\operatorname{SID}\rightarrow\mathcal{I}$ resolves each sequence to one or more items.

The retriever is inherently tied to the SID space through its token embeddings and output distributions. Let $M_{t_0}$ be the mapping used to label the training data and $M_{t_1}$ the mapping at inference time. A drift event occurs when $M_{t_0}\neq M_{t_1}$. We measure its extent over old items as
\begin{equation}
\gamma=\delta(M_{t_0},M_{t_1})=
\frac{1}{|\mathcal{I}|}\sum_{i\in\mathcal{I}}
\mathbb{I}\!\left[M_{t_0}^{-1}(i)\neq M_{t_1}^{-1}(i)\right].
\label{eq:drift}
\end{equation}
In the experiments, we denote the observed remapped fraction as $\gamma$ evaluated over the old-item set.


The central tension is stability versus freshness. Several routine operations can trigger mapping drift: refreshing value-based positions to track shifting market conditions, retraining an RQ-VAE or other item-to-SID clustering model after a large batch of new items alters the codebook, or more generally any procedure that reassigns existing items to different SIDs. Each of these invalidates part of the mapping learned by $f_\theta$. Unlike flat class labels, each SID in a generative retrieval model corresponds to a sequence of autoregressive decoding decisions: the model first chooses $s_1$, then $s_2$ conditioned on $s_1$, and so on until the full identifier is produced. Every Item-SID association is therefore not a single label but a learned decoding path through a tree-structured output space. When a mapping update reassigns items to different SIDs, it does not merely relabel training data; it redistributes output probability across previously learned paths, potentially invalidating entire branches of the decoding tree. \emph{Topology preservation}, keeping item-to-SID assignments unchanged, ensures that these learned paths remain valid. A wide beam may partially compensate for drift by exploring additional branches, whereas a narrow beam commits earlier and is more vulnerable.

\subsection{SID Construction}
The first two tokens $(s_1,s_2)$ form a semantic prefix obtained by quantizing item-content embeddings with an RQ-VAE trained under the CQ-SID query-item (QI) contrastive framework~\cite{zhu2026ecommerce}. The QI objective augments the standard VQ-VAE reconstruction and commitment losses with a bidirectional InfoNCE term that aligns item-content and query representations in a shared embedding space, so that the resulting codebook captures search-relevant semantic clusters rather than content-only neighborhoods. The codebook size is set to 8192x8192. 

Within each semantic prefix, the third token $s_3$ encodes a business-value prior. Following~\cite{ling2026businessvalue}, we compute
\begin{equation}
v(i)=\frac{0.1\cdot\mathrm{Click}(i)+0.3\cdot\mathrm{Cart}(i)+1.0\cdot\mathrm{Buy}(i)}{\mathrm{Exp}(i)+\epsilon},\quad\epsilon=10,
\label{eq:bizvalue}
\end{equation}
where $\mathrm{Exp}(i)$ denotes the exposure count of item~$i$, and $\mathrm{Click}$, $\mathrm{Cart}$, and $\mathrm{Buy}$ denote click, add-to-cart and purchase counts, respectively. Items within each prefix are ranked by $v(i)$, so that the third-level token directly reflects commercial importance.

\paragraph{Hybrid collision-aware assignment.}
In a collection of production items with continuous ingestion, the number of items under each semantic prefix grows monotonically over time. Maintaining a bijective item-to-token mapping in a bounded vocabulary is therefore infeasible: as new items accumulate, collisions (multiple items sharing the same identifier) become inevitable regardless of the initial codebook size. Rather than attempting to eliminate all collisions through an ever-expanding vocabulary, we adopt a hybrid collision-aware (HC) design that concentrates exact resolution where it matters most. For each prefix $P=(s_1,s_2)$, items are sorted by $v(i)$ and the top $K$ items receive unique third-level tokens, $s_3(i_j)=j$ for $j\in\{1,\ldots,K\}$; remaining items share bucket identifiers in fixed-size groups. We set $K=100$ and allocate up to 1024 additional collision buckets for the tail. As a reference, the full-sparse (FS) design assigns one unique identifier per item, requiring up to 8192 third-level buckets. HC preserves exact resolution for the high-value head while gracefully accommodating the long tail.

\paragraph{Non-invasive new-item insertion.}
Once a new item accumulates sufficient exposure to become eligible for the item pool, it first receives a semantic prefix from the RQ-VAE quantizer and is then inserted into the value ranking of its corresponding cluster. The item is inserted directly at its value-rank position, regardless of whether that position falls in the top-$K$ head or the collision-bucketed tail. The top-$K$ dedicated assignment is maintained only in the training-time SID set; at insertion time, a new item landing in the head may share a previously dedicated token with an existing item, and no old-item SID is modified. This ensures that new-item insertion has zero impact on the existing SID space.

\subsection{Retriever Model}
\label{sec:retriever}
The retriever is built on Qwen2.5-0.5B~\cite{qwen2025qwen25technicalreport} and training follows the three-stage pipeline of CQ-SID~\cite{zhu2026ecommerce}:
\begin{enumerate}
  \item \textbf{Item$\rightarrow$SID.} The model learns to generate item SIDs conditioned on item content alone, establishing the basic mapping from content representations to the SID space.
  \item \textbf{Query$\rightarrow$SID.} Query text is incorporated as an input to train the model to generate target item SIDs. This is achieved by leveraging query-SID associations derived from the majority of users' query-item interaction sequences.
  \item \textbf{User+Query$\rightarrow$SID.} User profile and interaction history are prepended to the query, enabling personalized retrieval that conditions SID generation on both user context and query intent.
\end{enumerate}

\subsection{Update Policies}
Each update cycle poses two independent decisions: \emph{which} items receive new SIDs (the mapping policy) and \emph{whether} the model is retrained to accommodate the updated mapping (the training strategy). We deliberately focus on simple, interpretable options rather than sophisticated continual-learning methods, for two reasons. First, production systems favor reliable, debuggable solutions over algorithmic complexity. Second, because the autoregressive retriever learns query-to-SID-path associations, any mapping change structurally invalidates part of these learned associations; the mapping policy is therefore expected to be the dominant factor, while the training procedure plays a supporting role.

\paragraph{Mapping policies.}
We consider three policies that span the stability--freshness spectrum. \emph{Full remapping} (Full-Re) reassigns third-level tokens to all items within each prefix using the latest value-rank; a large fraction of terminal token assignments change. \emph{Frozen mapping} keeps every old item's SID unchanged; eligible new items receive an available unique or shared-bucket token under their RQ-VAE prefix without modifying any existing assignment. \emph{Thresholded remapping} interpolates between the two: an old item is remapped only when its rank displacement exceeds a threshold~$t$,
\begin{equation}
|\operatorname{rank}_{t_1}(i)-\operatorname{rank}_{t_0}(i)|>t.
\end{equation}
Varying $t$ from $0$ to $\infty$ traces a continuum from Full-Re to Frozen mapping, allowing us to study the dose--response relationship between the remapping fraction $\gamma$ and retrieval quality.

\paragraph{Training strategies.}
Given any mapping policy, the model either remains unchanged (\emph{no-retrain}), or is continued from the previous checkpoint (\emph{warm-start}) and trained on accumulated new query--SID pairs labeled with the updated mapping. Warm-start training allows the model to learn associations involving newly inserted items and to reinforce existing ones under the current mapping. During warm-start updates, only retriever model training stage~3 in Section~\ref{sec:retriever} is re-executed.

The two dimensions combine independently, allowing different mapping policies to be evaluated with or without warm-start training. Figure~\ref{fig:system} contrasts the two extremes: Full-Re (red path) rewrites the terminal token assignments consumed by the existing decoder, causing topology disruption and degrading both HitRate and MRR; Frozen mapping with warm-start training (green path) extends the mapping with new items only and leverages warm-start training to improve retrieval quality for both old and new items.

\begin{table*}[t]
\centering
\caption{Day1 comparison of FS and HC identifiers: HitRate, MRR, and resolved items per query.}
\label{tab:sidbase}
\small
\setlength{\tabcolsep}{6pt}
\begin{tabular}{l rrrrr c rrrr}
\toprule
 & \multicolumn{5}{c}{Retrieval Quality} & & \multicolumn{4}{c}{Resolved Items} \\
\cmidrule(lr){2-6} \cmidrule(lr){8-11}
SID & HR@1 & HR@10 & HR@100 & HR@1000 & MRR & & Avg@B1 & Med@B1 & Avg@B1000 & Med@B1000 \\
\midrule
FS & 0.0584 & 0.2409 & 0.5134 & 0.7175 & 0.1181 & & 1 & 1 & 925 & 966 \\
HC & \textbf{0.0589} & \textbf{0.2457} & \textbf{0.5215} & \textbf{0.7183} & \textbf{0.1199} & & 1 & 1 & 1200 & 986 \\
\bottomrule
\end{tabular}
\label{tab:recall}
\end{table*}

\section{Experiments}
\subsection{Experimental Setup}
\paragraph{Datasets}
All experiments use query--item interaction logs from a large-scale e-commerce search platform. The observation window spans 67 days. The first 30 days train the initial model, the following 30 days form the first update and evaluation cycle, and the final 7 days form a subsequent non-overlapping update cycle initialized from the Day-30 model. Training interactions are split strictly before their evaluation events. 

\paragraph{Implementation Details}
The full-sparse (FS) codebook has size $8192\times8192\times8192$ across the three SID levels; the hybrid collision-aware is $8192\times8192\times1124$. The RQ-VAE is trained with exponential moving average (EMA) updates and a codebook reset strategy to stabilize quantizer training. The LLM retriever is trained on approximately 90 million logs for the 30-day cycle and 20 million for the 7-day cycle. All training runs on 64 GPUs with batch size $64$ and gradient accumulation over $4$ steps (effective batch size $256$). The initial model is trained for $10000$ steps at learning rate $4\times10^{-5}$; warm-start updates use a lower learning rate of $1\times10^{-5}$, with $5000$ steps for the 30-day cycle and $1200$ steps for the 7-day cycle.

\paragraph{Evaluation}
For each query, beam search generates candidate SID sequences, the current map resolves them to items, and HitRate records whether the target item is present. We report $\mathrm{HR}@B$ for $B\in\{1,10,100,1000\}$ and $\mathrm{MRR}@1000$. Day-1 denotes the first day following a model trained on the prior 30 days of data, which corresponds to the 31st day of the overall cycle. Subsequent references such as Day-30 and Day-37 are calculated relative to this Day-1 baseline. The same SID-expansion and metric computation are used across temporal variants. The experiments are organized around SID design, mapping drift, change-rate sensitivity, model adaptation, weight mixing, and cycle-to-cycle reproducibility.

\subsection{Collisions Are Not the Bottleneck}
Before investigating mapping drift, we first verify that the SID design itself, specifically the controlled collisions inherent in production-scale systems, does not already dominate retrieval quality.

We compare the full-sparse (FS) and hybrid-collision (HC) designs using the Day-1 evaluation. As shown in Table~\ref{tab:sidbase}, HC consistently outperforms FS across all beam sizes, achieving HR@1 = 0.0589 (+0.86\%), HR@1000 = 0.7183 (+0.11\%), and MRR = 0.1199 (+1.52\%). The table further reports the number of resolved items per query. Together with the uniformly higher HitRate and MRR, these results suggest that the collision mechanism effectively concentrates probability mass on relevant SID paths, while the bucket structure for tail items recovers additional candidates once those paths are selected. Moreover, the median number of retrieved items for HC at beam 1 is identical to that of FS (both 1) and only marginally higher at beam 1000 (986 vs. 966, the quantity is less than 1,000 because certain retrieved SIDs have no associated items.), indicating that controllable collisions do not cause the model to indiscriminately favor large clusters.

\subsection{Wide Beams Mask Mapping Drift}
\label{Wide Beams Mask Mapping Drift}
Next, we keep the item set fixed and investigate whether updating each item’s third-level SID token (based on its business value observed on Day 30) affects retrieval performance. We decouple two independently refreshable components: \emph{Map}, the serving-time SID-to-item lookup table, and \emph{Feat}, the SID representations within the model’s input interaction sequences. In the result tables, a check mark indicates that a component is refreshed to its latest version, whereas a cross indicates that it remains frozen. 

\begin{table}[H]
\centering
\caption{Old-item metrics on Day30 without adding new items. $\checkmark$: refreshed; $\times$: frozen.}
\label{tab:drift}
\small
\setlength{\tabcolsep}{3.2pt}
\begin{tabular}{lccrrrrr}
\toprule
SID & Map & Feat & HR@1 & HR@10 & HR@100 & HR@1000 & MRR \\
\midrule
FS & $\times$ & $\times$ & \textbf{0.0585} & \textbf{0.2324} & \textbf{0.4927} & \textbf{0.7002} & \textbf{0.1158} \\
FS & $\checkmark$ & $\checkmark$ & 0.0319 & 0.1268 & 0.3262 & 0.6333 & 0.0647 \\
FS & $\checkmark$ & $\times$ & 0.0222 & 0.0982 & 0.2890 & 0.6242 & 0.0492 \\
\midrule
HC & $\times$ & $\times$ & \textbf{0.0584} & \textbf{0.2344} & \textbf{0.5005} & \textbf{0.7035} & \textbf{0.1164} \\
HC & $\checkmark$ & $\checkmark$ & 0.0340 & 0.1394 & 0.3535 & 0.6449 & 0.0706 \\
HC & $\checkmark$ & $\times$ & 0.0226 & 0.1056 & 0.3172 & 0.6373 & 0.0519 \\
\bottomrule
\end{tabular}
\end{table}

Table~\ref{tab:drift} compares the fully frozen baseline (where both the map and feature representations are kept static) against two partial-refresh configurations. Refreshing both components significantly degrades performance, with HC HR@1 dropping from 0.0584 to 0.0340 (-41.78\%) and MRR falling from 0.1164 to 0.0706 (-39.35\%).

Notably, refreshing only the map while keeping input-sequence features frozen is even more detrimental than refreshing both. The two configurations differ in where the inconsistency lies. When both are refreshed ($\checkmark$/$\checkmark$), the model's conditioning context and the resolution table are mutually consistent, so the freshly generated SIDs and the map that resolves them describe the same value-rank; the residual loss stems only from the mismatch against the model's learned weights. When only the map is refreshed ($\checkmark$/$\times$), the model still emits SIDs consistent with the stale input features, but the serving map now resolves those SIDs under the new value-rank—so a decoding path that was coherent under the old features is re-pointed to different items, compounding a feature–map mismatch on top of the weight–map mismatch. The additional degradation therefore isolates the cost of internal inconsistency between the input representation and the resolution table, and reinforces our central argument: it is the reassignment of learned paths, not the freshness of the value signal per se, that drives drift.


The same comparison reveals substantially smaller degradation at beam size 1000: the HR@1000 decreases are 8.33\% and 9.41\% for the two HC refresh configurations. Across FS and HC, the narrow-beam HR@1 degradation is 41.78--62.05\%, whereas the wide-beam degradation is only 8.33-10.85\%. MRR, which averages over all ranks within beam 1000, shows an intermediate degradation of 39.35--55.41\%, confirming that drift primarily damages high-confidence paths. The observations are consistent with wide beams recovering some stale paths through redundant exploration and thereby concealing part of the failure mode.

Since HC consistently outperforms FS, we adopt HC as the default SID design for all subsequent experiments. Having established that controlled collisions do not degrade retrieval quality, the remaining source of performance loss in dynamic settings is the mapping update itself, which we investigate next.

\subsection{Remapping Rate Predicts Degradation}
Before studying how remapping affects old items, we first confirm whether these two strategies will have different impacts on new project retrieval.

We examine new-item retrieval without model retraining. Table~\ref{tab:newitem} compares Full-Re (all items remapped by Day-30 value) with Frozen (old SIDs frozen, new items inserted by value). The two policies produce nearly identical metrics across all beam sizes and MRR, indicating that new-item retrieval is insensitive to how old-item SIDs are represented in the input features.

\begin{table}[H]
\centering
\caption{New-item HitRate and MRR (Day30).}
\label{tab:newitem}
\small
\setlength{\tabcolsep}{4pt}
\begin{tabular}{lrrrrr}
\toprule
Policy & HR@1 & HR@10 & HR@100 & HR@1000 & MRR \\
\midrule
Full-Re & 0.0187 & 0.0828 & 0.2499 & 0.5133 & 0.0417 \\
Frozen & 0.0180 & 0.0835 & 0.2507 & 0.5127 & 0.0415 \\
\bottomrule
\end{tabular}
\end{table}

We then vary threshold $t$ to control the remapping fraction $\gamma$ of the old items. The SID map and input sequence features are refreshed consistently. Table~\ref{tab:gamma} shows a strictly monotonic pattern: as $\gamma$ decreases from 72.75\% to 0\%, HitRate and MRR increase at every beam size. Even the moderate threshold $t=10$ remains 12.4\% below complete freezing at beam size 1. Within the tested policies, partial remapping does not yield a useful stability-freshness compromise. 

Under full remapping (Full-Re), all metrics degrade monotonically: HR@1 drops by 42.66\% and MRR by 40.34\%, while HR@1000 exhibits the smallest decline of only 8.35\%, consistent with the masking effect observed in Section ~\ref{Wide Beams Mask Mapping Drift}.

Notably, inserting new items with frozen old SIDs (Frozen, $\gamma=0$) raises old-item HR@1000 from 0.7035 (Table~\ref{tab:drift}, without new items) to 0.7043, confirming that adding new-item signals to the feature sequence benefits old-item retrieval as well. 

\begin{table}[H]
\centering
\caption{Old-item metrics versus the remapped fraction $\gamma$.}
\label{tab:gamma}
\small
\setlength{\tabcolsep}{3.5pt}
\begin{tabular}{lrrrrrr}
\toprule
Policy & $\gamma$ & HR@1 & HR@10 & HR@100 & HR@1000 & MRR \\
\midrule
Full-Re & 72.75\% & 0.0332 & 0.1368 & 0.3523 & 0.6455 & 0.0692 \\
$t=10$ & 29.08\% & 0.0507 & 0.1996 & 0.4332 & 0.6568 & 0.1004 \\
$t=100$ & 7.36\% & 0.0553 & 0.2196 & 0.4694 & 0.6712 & 0.1097 \\
$t=500$ & 1.65\% & 0.0557 & 0.2217 & 0.4749 & 0.6769 & 0.1106 \\
Frozen & 0\% & \textbf{0.0579} & \textbf{0.2343} & \textbf{0.5011} & \textbf{0.7043} & \textbf{0.1160} \\
\bottomrule
\end{tabular}
\end{table}

\begin{figure}[t]
\centering
\includegraphics[width=\columnwidth]{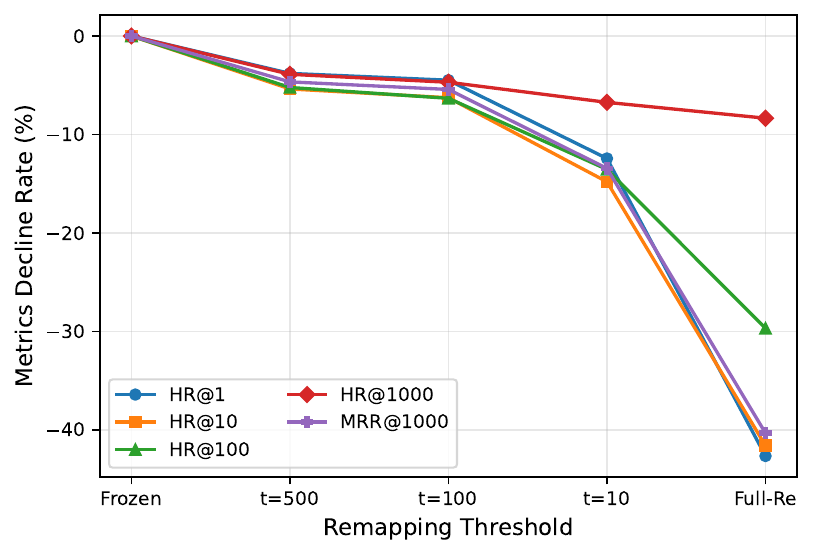}
\caption{\textbf{Monotonic remapping effect.} Metric decline rate (\%) relative to the frozen baseline as the remapping threshold becomes more aggressive. }
\label{fig:gamma}
\end{figure}

\subsection{Warm-Start Does Not Repair Remapping}
We now warm-start both Full-Re and Frozen from the same Day-1 model and train on the same accumulated Day1--30 data. Only the mapping policy differs. Table~\ref{tab:warmstart} shows higher observed HitRate and MRR for Frozen on both old and new items. At beam size 1, Frozen HitRate improves over Full-Re by 36.70\% on old items and 13.94\% on new items; MRR also shows a large gap (35.50\%/+9.56\% for old/new items). For old items, Full-Re after training still underperforms the no-retrain Frozen baseline at beam size 1000 (0.6923 vs. 0.7043 on Table~\ref{tab:gamma}) and on MRR (0.0941 vs. 0.1160).

Compared with the no-retrain policy, Frozen raises old-item HR@1000 from 0.7043 to 0.7272 and MRR from 0.1160 to 0.1276; new-item HR@1000 rises from 0.5127 to 0.6200 and MRR from 0.0415 to 0.0779. These observations are consistent with mapping preservation protecting established paths while warm-start training learns new Query-SID associations.

\begin{table}[H]
\centering
\caption{Warm-start comparison after the 30-day cycle.}
\label{tab:warmstart}
\small
\setlength{\tabcolsep}{3.5pt}
\begin{tabular}{llrrrrr}
\toprule
Items & Policy & HR@1 & HR@10 & HR@100 & HR@1000 & MRR \\
\midrule
\multirow{2}{*}{Old}
& Full-Re & 0.0466 & 0.1867 & 0.4309 & 0.6923 & 0.0941 \\
& Frozen & \textbf{0.0637} & \textbf{0.2597} & \textbf{0.5355} & \textbf{0.7272} & \textbf{0.1276} \\
\midrule
\multirow{2}{*}{New}
& Full-Re & 0.0330 & 0.1474 & 0.3684 & 0.6193 & 0.0711 \\
& Frozen & \textbf{0.0376} & \textbf{0.1590} & \textbf{0.3833} & \textbf{0.6200} & \textbf{0.0779} \\
\bottomrule
\end{tabular}
\end{table}

\begin{figure}[t]
\centering
\includegraphics[width=\columnwidth]{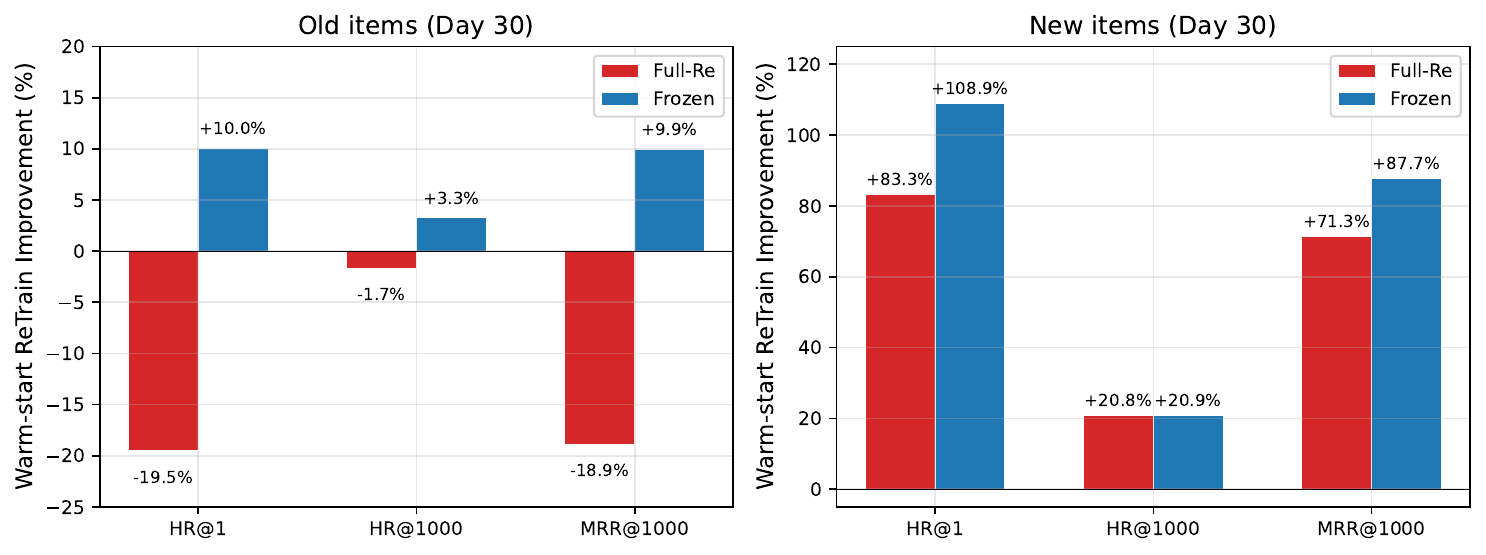}
\caption{\textbf{Warm-start training improvement over frozen baseline (Day-30).} Relative change (\%) for Full-Re and Frozen compared with the frozen (no-retraining) policy.}
\label{fig:warmstart}
\end{figure}

The results support a structural rather than purely optimization-based explanation. With Frozen and Full-Re, the initialization and accumulated training data are held fixed, and the mapping policy is the principal experimental difference. Frozen preserves established Query2SID associations, whereas Full-Re rewrites many SID assignments. A finite warm-start run must therefore relearn many old SID decisions while also learning new items. This interpretation is consistent with two empirical observations: degradation grows monotonically with $\gamma$, as expected if a larger remapped fraction invalidates more learned SID decisions; and the smaller degradation at a large beam is consistent with alternative-path exploration compensating for some mapping errors.

The effect of the warm-start training can be seen from Figure~\ref{fig:warmstart}. For old items (left), Full-Re degrades HR@1 by $-19.5\%$ and MRR by $-18.9\%$, while Frozen improves all three metrics. For new items (right), Frozen outperforms Full-Re on every metric, with HR@1 gaining $+108.9\%$ over frozen versus $+83.3\%$ for Full-Re (the no-retrain baseline corresponds to $\alpha=1.0$ in Table~\ref{tab:mixing}).

\paragraph{Scope of this claim.} In practical industrial environments, simple and reliable methods are generally preferred. Accordingly, our warm-start mechanism adopts a straightforward training strategy: it loads the parameters of the previously trained model and then fine-tunes them on the newly arrived data. Therefore, we do not claim that topological changes are inherently irreversible; a more robust continual learning framework may partially recover performance. Nevertheless, in the present context, preserving the original topology remains the optimal solution.

\subsection{No Benefit from Weight Mixing}
Warm-start training naturally raises a follow-up question at the parameter level. The previous checkpoint $\theta_{old}$ encodes stable associations for old items, whereas the warm-started checkpoint $\theta_{new}$ additionally encodes fresh associations for newly inserted items. A standard, training-free way to trade off this stability–plasticity tension is to interpolate the two checkpoints at inference time, in the spirit of model-soup~\cite{wortsman2022modelsoupsaveragingweights} and WiSE-FT-style weight averaging~\cite{wortsman2022robustfinetuningzeroshotmodels}; such interpolation is attractive in production because it needs no extra training and is tuned by a single scalar. We test whether weight mixing offers any additional benefit—that is, whether a useful middle ground exists at the parameter level. For Frozen policy, we interpolate the old and newly trained parameters at inference time:
\begin{equation}
\theta(\alpha)=\alpha\theta_{\mathrm{old}}+(1-\alpha)\theta_{\mathrm{new}}.
\end{equation}
Table~\ref{tab:mixing} reports the 30-day cycle. Old-item quality changes moderately across interpolation weights, whereas new-item quality depends strongly on the new checkpoint. A plausible explanation is that unchanged old-item SIDs are represented in both checkpoints, while new Query-SID associations appear only during the update; prediction-distribution analysis is needed to verify this hypothesis. The pure new checkpoint ($\alpha=0$) has the highest or near-highest observed value at every reported beam and on MRR for both item groups. At beam size 1, it raises new-item HitRate from 0.0180 to 0.0376 relative to the frozen old model, representing a more than twofold improvement. The tested interpolation settings provide no observed benefit.

\begin{table}[H]
\centering
\caption{Weight-mixing ablation for Frozen after 30 days.}
\label{tab:mixing}
\small
\setlength{\tabcolsep}{3.0pt}
\begin{tabular}{llrrrrr}
\toprule
Items & $\alpha$ & HR@1 & HR@10 & HR@100 & HR@1000 & MRR \\
\midrule
\multirow{4}{*}{Old}
& 0.0 & \textbf{0.0637} & \textbf{0.2597} & \textbf{0.5355} & \textbf{0.7272} & \textbf{0.1276} \\
& 0.1 & 0.0631 & 0.2590 & 0.5321 & 0.7270 & 0.1276 \\
& 0.5 & 0.0628 & 0.2502 & 0.5228 & 0.7193 & 0.1243 \\
& 1.0 & 0.0579 & 0.2343 & 0.5011 & 0.7043 & 0.1160 \\
\midrule
\multirow{4}{*}{New}
& 0.0 & \textbf{0.0376} & \textbf{0.1590} & \textbf{0.3833} & \textbf{0.6200} & \textbf{0.0779} \\
& 0.1 & 0.0339 & 0.1503 & 0.3744 & 0.6153 & 0.0727 \\
& 0.5 & 0.0233 & 0.1089 & 0.3141 & 0.5752 & 0.0534 \\
& 1.0 & 0.0180 & 0.0835 & 0.2507 & 0.5127 & 0.0415 \\
\bottomrule
\end{tabular}
\end{table}

Table~\ref{tab:mixing37} reports the same ablation in a subsequent seven-day cycle initialized from the Day-30 model and can be seen in Figure~\ref{fig:warmstart}. The qualitative pattern is similar: new-item quality depends strongly on the new checkpoint, while old-item quality is comparatively stable. At $\alpha=0.1$, a few old-item metrics (HR@10, HR@100, MRR) marginally exceed the $\alpha=0$ values. This suggests that, over a shorter cycle, retaining a small portion of the old model’s weights can slightly improve performance on older items. However, the gain is marginal and almost flat, likely because the data available within a short cycle is less diverse, whereas the old model has already learned from a large amount of historical data. Therefore, this limited improvement does not contradict the overall conclusion that directly using the new checkpoint is optimal or near-optimal.

\begin{table}[H]
\centering
\caption{Weight-mixing ablation for Frozen after an additional 7-day cycle (Day37).}
\label{tab:mixing37}
\small
\setlength{\tabcolsep}{3.0pt}
\begin{tabular}{llrrrrr}
\toprule
Items & $\alpha$ & HR@1 & HR@10 & HR@100 & HR@1000 & MRR \\
\midrule
\multirow{4}{*}{Old}
& 0.0 & \textbf{0.0687} & \underline{0.2754} & \underline{0.5490} & \underline{0.7328} & \textbf{0.1357} \\
& 0.1 & \underline{0.0686} & \textbf{0.2756} & \textbf{0.5491} & \textbf{0.7329} & \underline{0.1356} \\
& 0.5 & 0.0674 & 0.2714 & 0.5443 & 0.7306 & 0.1335 \\
& 0.9 & 0.0647 & 0.2631 & 0.5348 & 0.7248 & 0.1291 \\
\midrule
\multirow{4}{*}{New}
& 0.0 & \textbf{0.0492} & \textbf{0.2114} & \textbf{0.4468} & \textbf{0.6551} & \textbf{0.1021} \\
& 0.1 & 0.0466 & 0.2043 & 0.4434 & 0.6535 & 0.0983 \\
& 0.5 & 0.0341 & 0.1549 & 0.3795 & 0.6126 & 0.0747 \\
& 0.9 & 0.0301 & 0.1319 & 0.3269 & 0.5619 & 0.0644 \\
\bottomrule
\end{tabular}
\end{table}

\begin{figure}[t]
\centering
\includegraphics[width=\columnwidth]{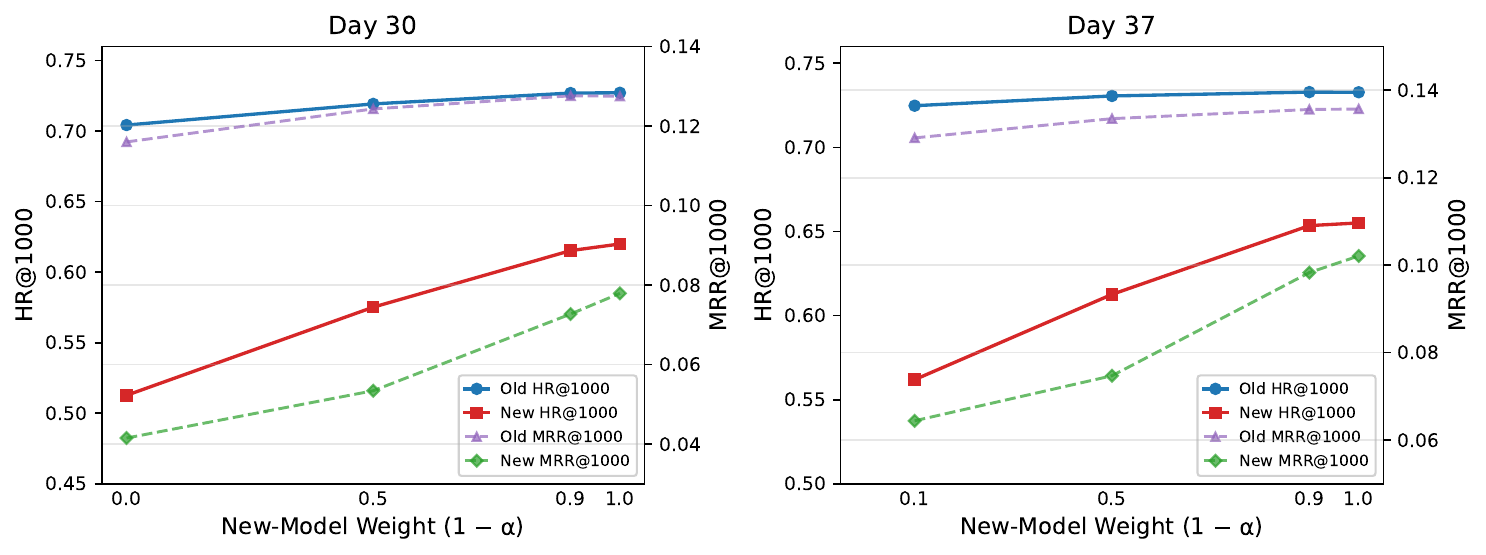}
\caption{\textbf{Asymmetric sensitivity to model mixing: HR@1000.} Day-30 (left) and Day-37 (right) weight-mixing ablations. The horizontal axis is the new-model weight $(1-\alpha)$. Solid lines show HR@1000 (left axis), dashed lines show MRR (right axis). Both metrics rise monotonically with the new-model weight, confirming that the pure new checkpoint ($1-\alpha=1$) is optimal or near-optimal.}
\label{fig:mixing}
\end{figure}

\subsection{Online A/B Test}
We deployed generative retrieval in the recall stage of the search system on TmallAPP, a major mobile e-commerce platform, where it works together with other recall channels. The generative retriever undergoes warm-start training on 7-days basis. The initial SIDs are constructed using a hybrid-collision method and updated daily. We adopt a topology-preserving policy that maintains the SIDs of existing items unchanged. For newly added items, the first two layers of the SID are generated via RQ-VAE, while the third layer is determined based on business value ranking. The system operates with dynamic beam sizes in [50, 200, 1000], maintaining an average end-to-end latency of approximately 50 ms. Compared with the baseline trained 2 months ago that freezes old items and inserts new items without cyclic training, a three-week online A/B test conducted on 20\% of live traffic validated the effectiveness of the proposed policies, demonstrating consistent gains in key metrics: UCTCVR increased by 0.29\% and GMV by 1.16\%.

\section{Discussion}
Our item SID deliberately embeds a time-varying business-value dimension (the third token $s_3$), which makes value-rank refresh a frequent and large-magnitude source of drift; in this sense the severity we report is characteristic of value-aware SID schemes. The underlying mechanism, however, is not specific to the value dimension: any operation that reassigns existing items to different SIDs—refreshing value positions, retraining the RQ-VAE codebook after a large ingestion batch, or de-duplicating collisions—invalidates part of the learned decoding paths, as formalized by $\delta(M_{t_0},M_{t_1})$ in Eq~\ref{eq:drift}. More generally, the drift magnitude is governed less by the presence of a value dimension per se than by how extensively the SID-assignment model is retrained: fully retraining the quantizer/codebook produces an entirely new SID space and hence the largest drift, whereas fine-tuning only a subset of parameters reassigns far fewer identifiers and yields proportionally smaller drift. Across this spectrum the direction of the effect and the preservation principle are unchanged. We therefore position topology preservation as a general operating principle whose urgency scales with how much of the mapping a given update reassigns.

The proposed policy, which involves freezing existing items, inserting new ones, and employing warm-starts, is intentionally designed for operational simplicity. We contend that this simplicity is a distinct advantage rather than a limitation. In production retrieval systems, update policies must inherently be reliable, easily debuggable, and broadly applicable across diverse item categories. Moreover, real-world deployments typically involve a temporal decoupling between serving model updates and item index table refreshes. During this transition period, any misalignment between the updating model and the item index can cause inference outputs to fail when mapping back to their corresponding items. This operational constraint also motivates our strategy of preserving the original SIDs for existing items. Therefore, the primary contribution of this work extends beyond the policy itself; it provides robust empirical evidence that such a straightforward approach is practically sufficient.

\section{Conclusion}
We investigate strategies for updating generative retrieval systems in e-commerce environments where items increase and business-value signals continuously evolve. Our experiments reveal that full identifier remapping induces severe semantic drift, whereas wide-beam evaluation demonstrates significantly less degradation. Across all tested update policies, both HitRate and MRR decrease monotonically as the proportion of remapped identifiers for existing items increases. Under identical warm-start conditions, retaining original identifiers for existing items while selectively inserting only eligible new ones yields higher HitRate than full remapping—a trend that remains consistent across subsequent dependent update cycles. Furthermore, models fine-tuned from previous checkpoints outperform naive model souping. These findings distill into a concise operational principle: insert new items but freeze existing ones; apply warm-starting when feasible, otherwise freeze the model; and strictly avoid altering the semantic identifiers (SIDs) the model has already learned. A three-week online A/B test on the mobile e-commerce platform TmallAPP search system validates this principle with UCTCVR +0.29\% and GMV +1.16\%.

\bibliographystyle{ACM-Reference-Format}
\balance
\bibliography{document}

@misc{qwen2025qwen25technicalreport,
      title={Qwen2.5 Technical Report}, 
      author={Qwen and : and An Yang and Baosong Yang and Beichen Zhang and Binyuan Hui and Bo Zheng and Bowen Yu and Chengyuan Li and Dayiheng Liu and Fei Huang and Haoran Wei and Huan Lin and Jian Yang and Jianhong Tu and Jianwei Zhang and Jianxin Yang and Jiaxi Yang and Jingren Zhou and Junyang Lin and Kai Dang and Keming Lu and Keqin Bao and Kexin Yang and Le Yu and Mei Li and Mingfeng Xue and Pei Zhang and Qin Zhu and Rui Men and Runji Lin and Tianhao Li and Tianyi Tang and Tingyu Xia and Xingzhang Ren and Xuancheng Ren and Yang Fan and Yang Su and Yichang Zhang and Yu Wan and Yuqiong Liu and Zeyu Cui and Zhenru Zhang and Zihan Qiu},
      year={2025},
      eprint={2412.15115},
      archivePrefix={arXiv},
      primaryClass={cs.CL},
      url={https://arxiv.org/abs/2412.15115}, 
}

@article{tay2022transformer,
  title={Transformer memory as a differentiable search index},
  author={Tay, Yi and Tran, Vinh and Dehghani, Mostafa and Ni, Jianmo and Bahri, Dara and Mehta, Harsh and Qin, Zhen and Hui, Kai and Zhao, Zhe and Gupta, Jai and others},
  journal={Advances in neural information processing systems},
  volume={35},
  pages={21831--21843},
  year={2022}
}

@article{rajput2024tiger,
  title={Recommender systems with generative retrieval},
  author={Rajput, Shashank and Mehta, Nikhil and Singh, Anima and Hulikal Keshavan, Raghunandan and Vu, Trung and Heldt, Lukasz and Hong, Lichan and Tay, Yi and Tran, Vinh and Samost, Jonah and others},
  journal={Advances in Neural Information Processing Systems},
  volume={36},
  pages={10299--10315},
  year={2023}
}

@article{deng2025onerec,
  title={Onerec: Unifying retrieve and rank with generative recommender and iterative preference alignment},
  author={Deng, Jiaxin and Wang, Shiyao and Cai, Kuo and Ren, Lejian and Hu, Qigen and Ding, Weifeng and Luo, Qiang and Zhou, Guorui},
  journal={arXiv preprint arXiv:2502.18965},
  year={2025}
}

@article{chen2025onesearch,
  title={Onesearch: A preliminary exploration of the unified end-to-end generative framework for e-commerce search},
  author={Chen, Ben and Guo, Xian and Wang, Siyuan and Liang, Zihan and Lv, Yue and Ma, Yufei and Xiao, Xinlong and Xue, Bowen and Zhang, Xuxin and Yang, Ying and others},
  journal={arXiv preprint arXiv:2509.03236},
  year={2025}
}

@misc{zhu2026ecommerce,
      title={Efficient Generative Retrieval for E-commerce Search with Semantic Cluster IDs and Expert-Guided RL}, 
      author={Jianbo Zhu and Xing Fang and Jing Wang and Mingmin Jin and Bokang Wang and Guangxin Song and Zhenyu Xie and Junjie Bai},
      year={2026},
      eprint={2605.14434},
      archivePrefix={arXiv},
      primaryClass={cs.IR},
      url={https://arxiv.org/abs/2605.14434}, 
}

@misc{ling2026businessvalue,
      title={Beyond Semantic IDs: Encoding Business-Value Ranking into Document Identifiers for Generative Retrieval}, 
      author={Gui Ling and Zhihong Chen and Yu Li and Tong Xiong and Kunhai Lin and Kaixuan Zhang and Yuliang Yan and Dan Ou and Haihong Tang and Bo Zheng},
      year={2026},
      eprint={2607.11392},
      archivePrefix={arXiv},
      primaryClass={cs.IR},
      url={https://arxiv.org/abs/2607.11392}, 
}

@misc{decao2021genre,
      title={Autoregressive Entity Retrieval}, 
      author={Nicola De Cao and Gautier Izacard and Sebastian Riedel and Fabio Petroni},
      year={2021},
      eprint={2010.00904},
      archivePrefix={arXiv},
      primaryClass={cs.CL},
      url={https://arxiv.org/abs/2010.00904}, 
}

@misc{damico2025spotify,
      title={Deploying Semantic ID-based Generative Retrieval for Large-Scale Podcast Discovery at Spotify}, 
      author={Edoardo D'Amico and Marco De Nadai and Praveen Chandar and Divita Vohra and Shawn Lin and Max Lefarov and Paul Gigioli and Gustavo Penha and Ilya Kopysitsky and Ivo Joel Senese and Darren Mei and Francesco Fabbri and Oguz Semerci and Yu Zhao and Vincent Tang and Brian St. Thomas and Alexandra Ranieri and Matthew N. K. Smith and Aaron Bernkopf and Bryan Leung and Ghazal Fazelnia and Mark VanMiddlesworth and Timothy Christopher Heath and Petter Pehrson Skiden and Alice Y. Wang and Doug J. Cole and Andreas Damianou and Maya Hristakeva and Reid Wilbur and Tarun Chillara and Vladan Radosavljevic and Pooja Chitkara and Sainath Adapa and Juan Elenter and Bernd Huber and Jacqueline Wood and Saaketh Vedantam and Jan Stypka and Sandeep Ghael and Martin D. Gould and David Murgatroyd and Yves Raimond and Mounia Lalmas and Paul N. Bennett},
      year={2026},
      eprint={2603.17540},
      archivePrefix={arXiv},
      primaryClass={cs.IR},
      url={https://arxiv.org/abs/2603.17540}, 
}

@misc{singh2024semanticids,
      title={Better Generalization with Semantic IDs: A Case Study in Ranking for Recommendations}, 
      author={Anima Singh and Trung Vu and Nikhil Mehta and Raghunandan Keshavan and Maheswaran Sathiamoorthy and Yilin Zheng and Lichan Hong and Lukasz Heldt and Li Wei and Devansh Tandon and Ed H. Chi and Xinyang Yi},
      year={2024},
      eprint={2306.08121},
      archivePrefix={arXiv},
      primaryClass={cs.IR},
      url={https://arxiv.org/abs/2306.08121}, 
}

@misc{zhan2026tsgrtaobaosearchgenerative,
      title={TSGR: Taobao Search Generative Retrieval}, 
      author={Tianyu Zhan and Gui Ling and Tong Xiong and Kunhai Lin and Yang Wang and Kaixuan Zhang and Zhihong Chen and Yuliang Yan and Dan Ou and Shengyu Zhang and Haihong Tang and Bo Zheng},
      year={2026},
      eprint={2607.18796},
      archivePrefix={arXiv},
      primaryClass={cs.IR},
      url={https://arxiv.org/abs/2607.18796}, 
}

@misc{zheng2025mtgrec,
      title={Pre-training Generative Recommender with Multi-Identifier Item Tokenization}, 
      author={Bowen Zheng and Enze Liu and Zhongfu Chen and Zhongrui Ma and Yue Wang and Wayne Xin Zhao and Ji-Rong Wen},
      year={2025},
      eprint={2504.04400},
      archivePrefix={arXiv},
      primaryClass={cs.IR},
      url={https://arxiv.org/abs/2504.04400}, 
}

@misc{chen2024selfimproving,
      title={Enhancing Item Tokenization for Generative Recommendation through Self-Improvement}, 
      author={Runjin Chen and Mingxuan Ju and Ngoc Bui and Dimosthenis Antypas and Stanley Cai and Xiaopeng Wu and Leonardo Neves and Zhangyang Wang and Neil Shah and Tong Zhao},
      year={2024},
      eprint={2412.17171},
      archivePrefix={arXiv},
      primaryClass={cs.LG},
      url={https://arxiv.org/abs/2412.17171}, 
}

@misc{liu2026topology,
      title={Topology-Aware Tokenization for Generative Recommendation}, 
      author={Yaokun Liu and Yifan Liu and Zhenrui Yue and Gyuseok Lee and Zelin Li and Ruichen Yao and Dong Wang},
      year={2026},
      eprint={2607.18600},
      archivePrefix={arXiv},
      primaryClass={cs.IR},
      url={https://arxiv.org/abs/2607.18600}, 
}

@misc{hu2025quasid,
      title={Stop Treating Collisions Equally: Qualification-Aware Semantic ID Learning for Recommendation at Industrial Scale}, 
      author={Zheng Hu and Yuxin Chen and Yongsen Pan and Xu Yuan and Yuting Yin and Daoyuan Wang and Boyang Xia and Zefei Luo and Hongyang Wang and Songhao Ni and Dongxu Liang and Jun Wang and Shimin Cai and Tao Zhou and Fuji Ren and Wenwu Ou},
      year={2026},
      eprint={2603.00632},
      archivePrefix={arXiv},
      primaryClass={cs.IR},
      url={https://arxiv.org/abs/2603.00632}, 
}

@misc{wang2026topogr,
      title={TopoGR: Revealing and Preserving Latent Structure of Semantic ID in Generative Recommendation}, 
      author={Ziyu Zheng and Zhengshun Du and Yaming Yang and Bin Tong and Guan Wang and Meng Yan and Ziyu Guan and Wei Zhao},
      year={2026},
      eprint={2607.25216},
      archivePrefix={arXiv},
      primaryClass={cs.IR},
      url={https://arxiv.org/abs/2607.25216}, 
}

@misc{dong2025h2rec,
      title={The Best of the Two Worlds: Harmonizing Semantic and Hash IDs for Sequential Recommendation}, 
      author={Ziwei Liu and Yejing Wang and Wanyu Wang and Wang Zejian and Qidong Liu and Zijian Zhang and Chong Chen and Wei Huang and Xiangyu Zhao},
      year={2026},
      eprint={2512.10388},
      archivePrefix={arXiv},
      primaryClass={cs.IR},
      url={https://arxiv.org/abs/2512.10388}, 
}

@inproceedings{tang2026static,
   title={Vectorizing the Trie: Efficient Constrained Decoding for LLM-based Generative Retrieval on Accelerators},
   url={http://dx.doi.org/10.1145/3770855.3818506},
   DOI={10.1145/3770855.3818506},
   booktitle={Proceedings of the 32nd ACM SIGKDD Conference on Knowledge Discovery and Data Mining V.2},
   publisher={ACM},
   author={Su, Zhengyang and Katsman, Isay and Wang, Yueqi and He, Ruining and Heldt, Lukasz and Keshavan, Raghunandan and Wang, Shao-Chuan and Yi, Xinyang and Gao, Mingyan and Dalal, Onkar and Hong, Lichan and Chi, Ed H. and Han, Ningren},
   year={2026},
   month=Aug, pages={8009–8020} }

@misc{fu2025forge,
      title={FORGE: Forming Semantic Identifiers for Generative Retrieval in Industrial Datasets}, 
      author={Kairui Fu and Tao Zhang and Shuwen Xiao and Ziyang Wang and Xinming Zhang and Chenchi Zhang and Yuliang Yan and Junjun Zheng and Xiangheng Kong and Shengyu Zhang and Kun Kuang and Yuning Jiang},
      year={2026},
      eprint={2509.20904},
      archivePrefix={arXiv},
      primaryClass={cs.IR},
      url={https://arxiv.org/abs/2509.20904}, 
}

@misc{guan2026dream,
      title={DREAM: Dynamic Refinement of Early Assignment Mappings}, 
      author={Liwei Guan and Huanjie Wang and Hongwei Zhang and Linxun Chen and Zhaojie Liu},
      year={2026},
      eprint={2606.06947},
      archivePrefix={arXiv},
      primaryClass={cs.IR},
      url={https://arxiv.org/abs/2606.06947}, 
}

@misc{mehta2023dsiplus,
      title={DSI++: Updating Transformer Memory with New Documents}, 
      author={Sanket Vaibhav Mehta and Jai Gupta and Yi Tay and Mostafa Dehghani and Vinh Q. Tran and Jinfeng Rao and Marc Najork and Emma Strubell and Donald Metzler},
      year={2023},
      eprint={2212.09744},
      archivePrefix={arXiv},
      primaryClass={cs.CL},
      url={https://arxiv.org/abs/2212.09744}, 
}

@misc{huynh2024promptdsi,
      title={PromptDSI: Prompt-based Rehearsal-free Continual Learning for Document Retrieval}, 
      author={Tuan-Luc Huynh and Thuy-Trang Vu and Weiqing Wang and Yinwei Wei and Trung Le and Dragan Gasevic and Yuan-Fang Li and Thanh-Toan Do},
      year={2025},
      eprint={2406.12593},
      archivePrefix={arXiv},
      primaryClass={cs.IR},
      url={https://arxiv.org/abs/2406.12593}, 
}

@inproceedings{chen2023clever,
   title={Continual Learning for Generative Retrieval over Dynamic Corpora},
   url={http://dx.doi.org/10.1145/3583780.3614821},
   DOI={10.1145/3583780.3614821},
   booktitle={Proceedings of the 32nd ACM International Conference on Information and Knowledge Management},
   publisher={ACM},
   author={Chen, Jiangui and Zhang, Ruqing and Guo, Jiafeng and de Rijke, Maarten and Chen, Wei and Fan, Yixing and Cheng, Xueqi},
   year={2023},
   month=Oct, pages={306–315},
   collection={CIKM ’23} }

@inproceedings{baikalov2026staleness,
   title={Mitigating Collaborative Semantic ID Staleness in Generative Retrieval},
   url={http://dx.doi.org/10.1145/3805712.3809877},
   DOI={10.1145/3805712.3809877},
   booktitle={Proceedings of the 49th International ACM SIGIR Conference on Research and Development in Information Retrieval},
   publisher={ACM},
   author={Baikalov, Vladimir and Bagautdinov, Iskander and Muravyov, Sergey},
   year={2026},
   month=July, pages={3602–3608} }

@misc{feng2026dact,
      title={Drift-Aware Continual Tokenization for Generative Recommendation}, 
      author={Yuebo Feng and Jiahao Liu and Mingzhe Han and Dongsheng Li and Hansu Gu and Peng Zhang and Tun Lu and Ning Gu},
      year={2026},
      eprint={2603.29705},
      archivePrefix={arXiv},
      primaryClass={cs.IR},
      url={https://arxiv.org/abs/2603.29705}, 
}

@misc{den2026icicle,
      title={ICICLE: Expanding Retrieval with In-Context Documents}, 
      author={Yu-Chen Den and Yung-Yu Shih and Zhi Rui Tam and Kuan-Yu Chen and Pu-Jen Cheng and Yun-Nung Chen and Eugene Yang},
      year={2026},
      eprint={2605.26902},
      archivePrefix={arXiv},
      primaryClass={cs.IR},
      url={https://arxiv.org/abs/2605.26902}, 
}

@inproceedings{mekonnen2026pamt,
   title={A Parametric Memory Head for Continual Generative Retrieval},
   url={http://dx.doi.org/10.1145/3805712.3809725},
   DOI={10.1145/3805712.3809725},
   booktitle={Proceedings of the 49th International ACM SIGIR Conference on Research and Development in Information Retrieval},
   publisher={ACM},
   author={Mekonnen, Kidist Amde and Tang, Yubao and de Rijke, Maarten},
   year={2026},
   month=July, pages={1347–1358} }

@misc{huynh2025mixlora,
      title={MixLoRA-DSI: Dynamically Expandable Mixture-of-LoRA Experts for Rehearsal-Free Generative Retrieval over Dynamic Corpora}, 
      author={Tuan-Luc Huynh and Thuy-Trang Vu and Weiqing Wang and Trung Le and Dragan Gašević and Yuan-Fang Li and Thanh-Toan Do},
      year={2026},
      eprint={2507.09924},
      archivePrefix={arXiv},
      primaryClass={cs.IR},
      url={https://arxiv.org/abs/2507.09924}, 
}

@misc{wortsman2022modelsoupsaveragingweights,
      title={Model soups: averaging weights of multiple fine-tuned models improves accuracy without increasing inference time}, 
      author={Mitchell Wortsman and Gabriel Ilharco and Samir Yitzhak Gadre and Rebecca Roelofs and Raphael Gontijo-Lopes and Ari S. Morcos and Hongseok Namkoong and Ali Farhadi and Yair Carmon and Simon Kornblith and Ludwig Schmidt},
      year={2022},
      eprint={2203.05482},
      archivePrefix={arXiv},
      primaryClass={cs.LG},
      url={https://arxiv.org/abs/2203.05482}, 
}

@misc{wortsman2022robustfinetuningzeroshotmodels,
      title={Robust fine-tuning of zero-shot models}, 
      author={Mitchell Wortsman and Gabriel Ilharco and Jong Wook Kim and Mike Li and Simon Kornblith and Rebecca Roelofs and Raphael Gontijo-Lopes and Hannaneh Hajishirzi and Ali Farhadi and Hongseok Namkoong and Ludwig Schmidt},
      year={2022},
      eprint={2109.01903},
      archivePrefix={arXiv},
      primaryClass={cs.CV},
      url={https://arxiv.org/abs/2109.01903}, 
}

\appendix

\end{document}